\documentclass[12pt]{article}
\usepackage{graphicx}

\newcommand\pubnumber{SNSN-323-63}
\newcommand\pubdate{\today}

\def\napoli{Physics Department, Brookhaven National Laboratory, Upton, NY 11973, USA}
\def\support{
}
\def\Title#1{\begin{center} {\Large #1 } \end{center}}
\def\Author#1{\begin{center}{ \sc #1} \end{center}}
\def\Address#1{\begin{center}{ \it #1} \end{center}}

\newcommand\pubblock{\rightline{\begin{tabular}{l} \pubnumber\\
         \pubdate  \end{tabular}}}
\newenvironment{Abstract}{\begin{quotation}  }{\end{quotation}}
\newenvironment{Presented}{\begin{quotation} \begin{center} 
             PRESENTED AT\end{center}\bigskip 
      \begin{center}\begin{large}}{\end{large}\end{center} \end{quotation}}
\def\Acknowledgements{\bigskip  \bigskip \begin{center} \begin{large}
             \bf ACKNOWLEDGEMENTS \end{large}\end{center}}

\def\beq{\begin{equation}}
\def\eeq#1{\label{#1}\end{equation}}
\def\eeqn{\end{equation}}

\def\beqa{\begin{eqnarray}}
\def\eeqa#1{\label{#1}\end{eqnarray}}
\def\eeqan{\end{eqnarray}}

\let\bar=\overbar

\def\Dslash{\not{\hbox{\kern-4pt $D$}}}
\def\dslash{\not{\hbox{\kern-2pt $\del$}}}

\def\msb{{\bar{\ssstyle M \kern -1pt S}}}

\begin{document}
\begin{titlepage}
\pubblock

\vfill
\Title{Recent progress in Nuclear Parton Distributions}
\vfill
\Author{P. Zurita\support}
\Address{\napoli}
\vfill
\begin{Abstract}
The determination of the parton distribution functions (PDFs) is crucial for a complete understanding of the protons and neutrons that make most of the visible matter in the universe. Years of dedicated studies have yielded a quite precise knowledge of the behavior of partons moving collinearly within a proton. However Deep Inelastic Scattering (DIS) experiments off nuclei have shown a non-trivial difference with respect to DIS in protons, hinting that the partons in a nuclear medium do not behave the same way as in a free proton. In this work we will discuss the latest results in nuclear parton distribution functions (nPDFs) and how data from planned future experiments can help broaden our understanding of the nPDFs.
\end{Abstract}
\vfill
\begin{Presented}
The Thirteenth Conference on the Intersections of Particle and Nuclear Physics (CIPANP 2018), May 29 - June 3, 2018 Palm Springs, CA.
\end{Presented}
\vfill
\end{titlepage}
\def\thefootnote{\fnsymbol{footnote}}
\setcounter{footnote}{0}

\section{Introduction}

The accurate description of the partonic behaviour is a fundamental piece for understanding the world around us. The simplest picture of the proton is given by the collinear factorised PDFs, where the initial state is described as a collection of partons, each carrying a fraction $x$ of the momentum of the proton. How many partons can be seen also depends on some characteristic scale of the process (usually called $Q^{2}$), typically required to be above a few GeV$^{2}$ in order for the theory to lie in the perturbative domain. Decades of theoretical and experimental efforts have provided a very sophisticated description of the PDFs in the free proton \cite{Harland-Lang:2014zoa,Ball:2017nwa,Abdolmaleki:2018jln,Hou:2017khm}. 

However the same can not be said about the nPDFs. DIS experiments off nuclei in the early 80's have shown that the nuclear medium affects non trivially the known proton PDFs \cite{Aubert:1981gv}. Four options open then: a) the fundamental interactions are the same but the PDFs have to be different, b) the fundamental interactions are different in the medium and the partonic cross-section have to be recomputed while the PDFs remain the same, c) both, and d) the factorisation picture is no longer valid. While in principle all these scenarios are equally possible, the simplest and most popular option is a). That is, assuming that one can still write the observables in a factorised form, and then extracting the nPDF from global fits to the world data. Of course there is no warranty of this being the right option, but almost 20 years of intense work has provided no evidence against this approach. On the contrary, nPDFs are determined each year with higher and higher accuracy and using more refined techniques as new data are published \cite{Eskola:2016oht,Kovarik:2015cma,Khanpour:2016pph}.  

Moreover nPDFs are not only important on their own right. On the one hand nuclear experiments are relevant for the determination of the free proton PDFs, as neutral current (NC) DIS off deuteron and charged current (CC) DIS of iron ($Fe$) and lead ($Pb$) are used for the flavour separation of the partonic species. On the other hand they are key for understanding the background for new phenomena in heavy-ion collisions. It is therefore not surprising then that current and future experiments include nuclear studies in their programs. This document is organised as follows: in the Sec.\ref{EPPS16} main points of the latest nPDF analysis are presented, and Sec. \ref{future} discuss some of the improvements that can be achieved for the  nPDFs in future experiments. Finally Sec. \ref{summary} summarises the current status of nPDFs.
 
\section{EPPS16}\label{EPPS16}

\begin{figure}[htb]
\centering
\includegraphics[height=3.in]{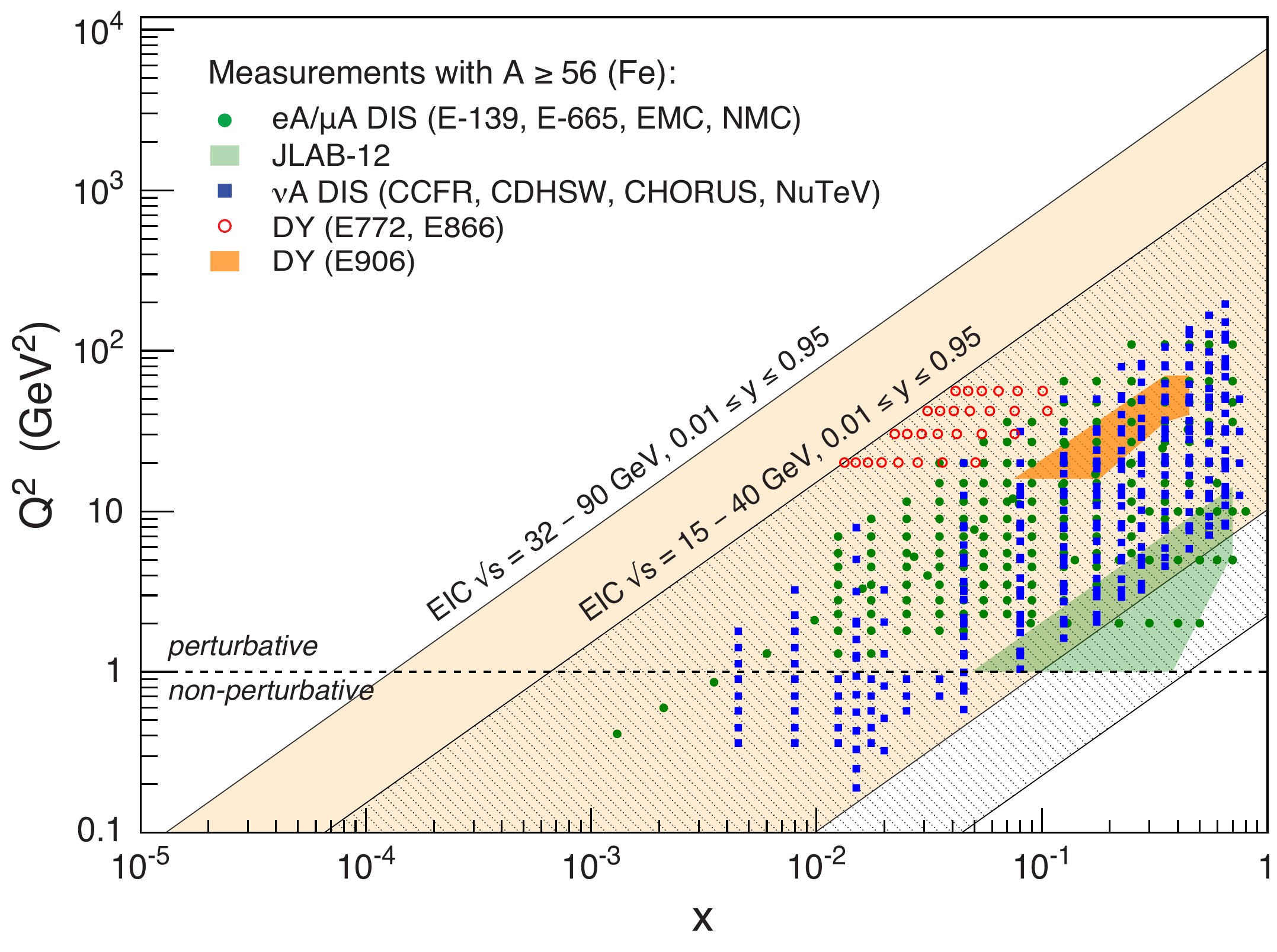}
\caption{Kinematic coverage of DIS and Drell-Yan experiments with nuclei used in the determination of nPDFs. The shaded bands correspond to the Electron-Ion Collider expected coverage (see text for details). Here there are not included the areas from electroweak bosons and di-jets measured at the LHC in $p+Pb$ collisions, which can be found in \cite{Eskola:2016oht}. Plot taken from \cite{Aschenauer:2017oxs}.}
\label{fig:eic-kin}
\end{figure}

The determination of the nPDFs is done by the same procedure as those for the free proton, i.e. by performing a global fit to the world data, but in this case the parametrizations include some $A$ dependence. Several sets have been extracted in the last two decades with different parametrizations and using diverse strategies, giving all of them adequate descriptions of the data. However, despite all the effort put in their determination, the nPDFs are not well constrained, particularly in the low $x$ region, due to experiments not spanning a broad enough region of the kinematic space. DIS experiments off nuclei are typically limited to the $x > 10^{-2}$ and $Q^{2} < 100$ GeV$^{2}$ region shown by dots and squares in Fig. \ref{fig:eic-kin}. 
This is at least two decades less in $x$ and almost three orders of magnitude less in $Q^{2}$ than the HERA collider. Any prediction beyond this region comes from an extrapolation tightly linked to the initial functional form and therefore conclusions must be extracted carefully. 

The latest global analysis published is EPPS16 \cite{Eskola:2016oht}, which includes data from NC and CC DIS with fixed targets, Drell-Yan processes, pion production at RHIC and, for the first time, data measured at the LHC in $p+Pb$. While the latter help broaden the kinematic space, they are not decisive in the full fit, the bulk of the data coming from NC and CC. A very important result is the fact that including the CC data seems to introduce no extra tension, as already noticed in \cite{deFlorian:2011fp}, which supports the idea that nuclear effects are universal {\footnote{This is still not a closed issue, as CC experiments have explored so far a quite restricted kinematic region.}}. This allowed the authors to attempt a full separation of the nuclear effects for each partonic species. The functional forms for the ratio between the nuclear and the proton PDFs at the initial scale $Q_{0}=1.3$ GeV are given by
\begin{equation}
R_{\rm EPPS16}(x) = 
\left\{
\begin{array}{lc}
a_0 + a_1(x-x_a)^2 & x \leq x_a \\
b_0 + b_1x^\alpha + b_2x^{2\alpha} + b_3x^{3\alpha} & x_a \leq x \leq x_e \\
c_0 + \left(c_1-c_2x \right) \left(1-x\right)^{-\beta} & x_e \leq x \leq 1.
\end{array}
\right. \label{EPPS16R}
\end{equation}
with one $R$ for each valence and sea quark, and one for the gluon. Imposing charge and momentum conservation, and merging some parameters due to lack of sensitivity of the data, the study increased the number of parameters from $15$ in their previous analysis \cite{Eskola:2009uj} to $20$. This has an impact on the theoretical uncertainty bands, making them more broad than in \cite{Eskola:2009uj} as a consequence of the data not having enough constraining power.





\section{Future experiments}\label{future}
It is clear that precise results covering a wider region of the kinematic space are needed in order to improve our understanding of the nPDFs. In this section we will briefly discuss some of the proposed future experiments. While these have broad physics programs, in the following we will touch only upon some of the measurements directly concerning the nuclear gluon density. All impact studies shown here are based on pseudo data generated under specific assumptions and, in consequence, the results must be interpreted keeping them in mind. 

\subsection{The Electron-Ion Collider}

The Electron-Ion Collider (EIC) is a proposed facility to be built in the EE.UU., colliding both polarized and unpolarized electrons and protons/nuclei to investigate with great precision their inner structure \cite{Accardi:2012qut}. While the final parameters of the accelerator are yet to be decided, the expected kinematic coverage is given by the light orange and shaded bands in Fig. \ref{fig:eic-kin}.  

\begin{figure}[htb]
\centering
\includegraphics[height=3.in]{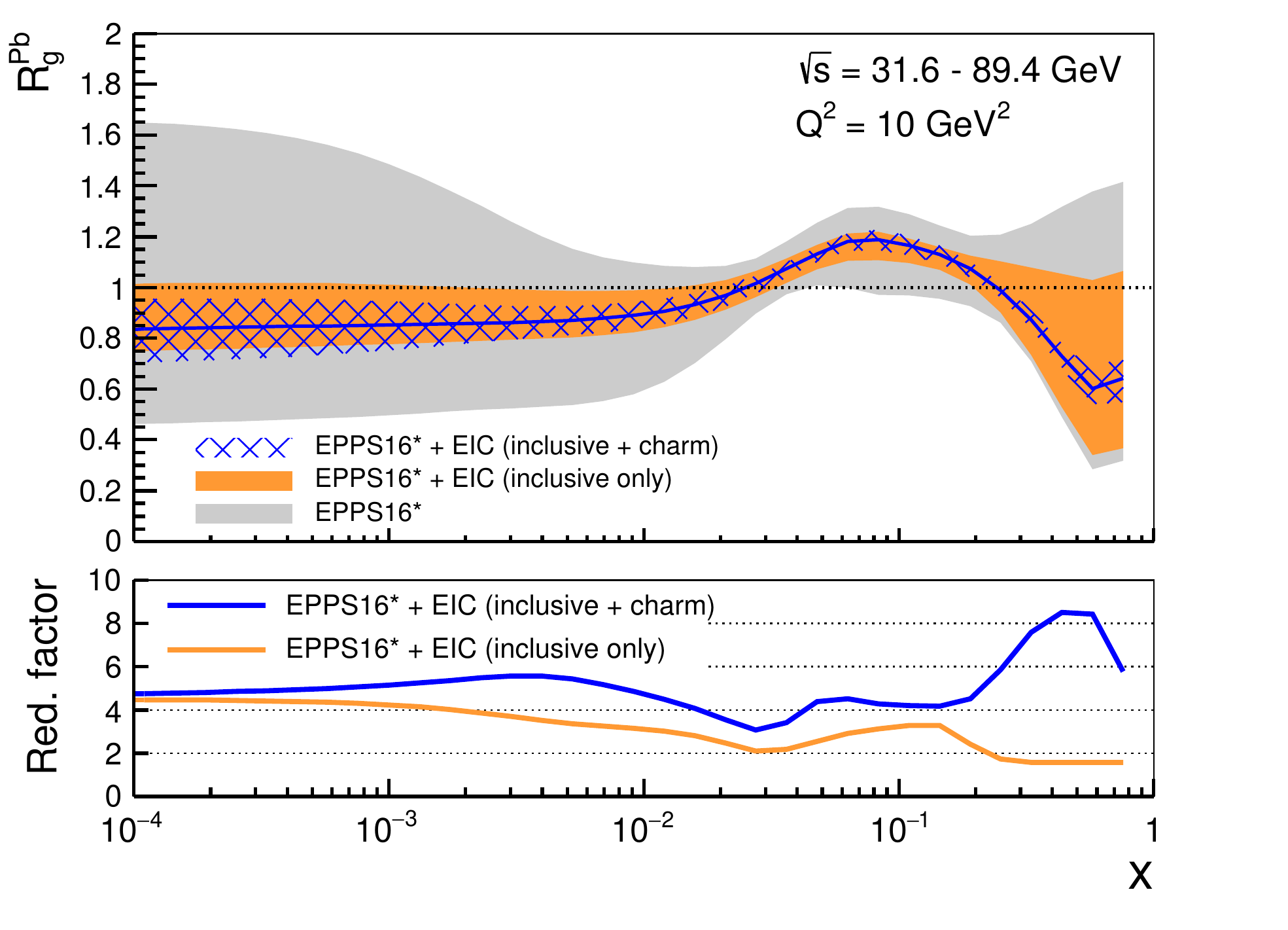}
\caption{Ratio of the nuclear gluon w.r.t. the free proton gluon at $Q^{2}=10$ GeV$^{2}$. The grey band corresponds to EPPS16*, the orange band to EPPS16* including inclusive EIC pseudo data, and the blue band to the inclusive plus charm pseudo data included in EPPS16*. Plot taken from \cite{Aschenauer:2017jsk}.}
\label{fig:eic-impact}
\end{figure}

As the core of a PDF analysis consist of cross-sections ($\sigma$), the first proposed observable is using the DIS inclusive reduced $\sigma$ ($\sigma_{r}$) {\footnote{The reduced cross-section is $\sigma$ divided by the Mott cross-section.}}. Furthermore one could explore the heavy-quarks associated $\sigma_{r}^{h\bar{h}}$, in particular the one for the charm quark which can be measured by tagging kaons in the final state. The full study of the impact for these pseudo data was presented in \cite{Aschenauer:2017oxs}, here we will highlight the results. The analysis consisted on including pseudo data for $\sigma_{r}$ and $\sigma_{r}^{c\bar{c}}$ into EPPS16* {\footnote {A re-fit of EPPS16 with the same data in it, but allowing for a more flexible functional form for the  low-$x$ gluon. For details about the parameterization and the fit results see \cite{Aschenauer:2017oxs}.}} and exploring how much and why the fitted parameters were modified. The result for a centre of mass energy (c.m.e.) range from $31.6$ to $89.4$ GeV is shown in Fig. \ref{fig:eic-impact}. There we plot the ratios of $Pb$ to proton gluon densities as a function of $x$ for $Q^{2}=10$ GeV$^{2}$. The grey band represents the theoretical uncertainty of EPPS16*  and the orange band corresponds to repeating the fit that incorporates the inclusive pseudo data. As can be seen in the lower panel, the reduction factor for the gluon uncertainty amounts to almost a factor 4 in the low $x$ region. The blue dashed band is the result of adding also into the fit $\sigma_{r}^{c\bar{c}}$. This has negligible impact at low $x$, but an incredible relevance for the  high-$x$ region, where the reduction factor can be as high as 8, due to the $c\bar{c}$ pair being radiated by a high-$x$ gluon. 
A similar but less pronounce effect was observed when reducing the highest c.m.e.  

Other observables that have been proposed to constrain the gluon at the EIC are jets and di-jets \cite{Klasen:2017kwb,Klasen:2018gtb}. For the former, the gluon initiated processes give more than $60\%$ of the total $\sigma$ for $x  < 10^{-3}$ and are expected to improve the precision of the gluon density by a factor up to $5$, while for the latter the gluon contribution to $\sigma$ amounts from $30$ to $70\%$ for of the total cross-section for $x > 0.1$.

\subsection{The LHeC and AFTER@LHC}

The Large Hadron electron Collider (LHeC) is a proposal to incorporate an electron beam in the LHC tunnel \cite{AbelleiraFernandez:2012cc}. The project presents some similarities with the EIC, but due to its much higher c.m.e. and the impossibility to polarise the proton/nucleus beam, the corresponding physics programs present little overlap, enough to cross-check some results but not so that their simultaneous existence could be redundant. Detailed studies on future DIS experiments at the LHeC have been performed, with conclusions similar to the ones shown above for the EIC. For details we refer the reader to \cite{Helenius:2016hcu}.  

The project ``A Fixed-Target Programme at the LHC" (AFTER@LHC) proposes to use the LHC beams with fixed targets in order to study the high-$x$ region. With a forward coverage, easily exchangeable target, hight luminosity and the possibility of polarising the target, it aims to profit from the high energy beams to complement current and future studies, closing gaps in the explored kinematic space. For detailed impact studies on the possibilities offered by AFTER@LHC, see \cite{Hadjidakis:2018ifr}. 

\section{Summary}\label{summary}

The PDFs are a key piece in the description of initial state that lie at the foundation of the Standard Model and are the object of in depth studies. In recent years the interest in the medium modification of the initial state has increased in the QCD community, due to their tight link to the PDFs: one can not fully determine the PDFs without nPDFs, and viceversa. Nowadays several nPDFs sets, periodically updated, are available, all providing adequate descriptions of the data. However the degree of precision in the nuclear case is not comparable to the one found in the proton PDFs, a consequence of the lack of precise data and the limited coverage of the kinematic space. 

The present status can be drastically changed by using data from past, current and future experiments. In the latter case detailed studies are undergoing in order to provide not only a one dimensional picture of the nucleons, but to fully characterise all aspects of the partonic behaviour. 

\Acknowledgements

P. Z. acknowledges the support by the U.S. Department of Energy under Contract No. DE-SC0012704.

\end{document}